# L$_\alpha$ line of dark positronium as nongravitational detection of dark matter

[1] V. Burdyuzha, [2] V. Charugin

[1] Astro-Space Center, Lebedev Physical Institute, Russian Academy of Sciences,

Profsoyuznaya Str. 84/32, 117997 Moscow, Russian Federation

[2] Moscow Pedagogical State University, M. Pirogovskaya Str.1/1,

119991 Moscow, Russian Federation

An attempt to predict the new atomic dark matter lines is done on the example of a dark lepton atom - positronium. Its Layman-alpha line with the energy near 3 GeV may be observable if the appropriate conditions realize. For this we have studied a γ-ray excess in the Center of our Galaxy. In principle, this excess may be produced by the L$_\alpha$ line of a dark positronium in the medium with Compton scattering. The possibility of observations of an annihilation line (E~300 TeV) of dark positronium is also predicted. Other proposals to observe the atomic dark matter are shortly described. Besides, H$_\alpha$ line (1.3μ) of usual positronum must be observable in the direction on the Center of our Galaxy.

## Introduction

Dark matter (DM) makes up the majority of matter density in our Universe. Its ubiquitous abundance can be measured through its gravitational influence, but we know little about DM microphysical properties, such as the mass, spin, and any nongravitational interactions. On the other hand we have interest indications on the possible dark matter detection which have been recently found [1-3]. 3.55 keV X-ray emission line could not be identified with any known spectral line of baryon elements. It was tempting to speculate that it was the decay of a particle contributing all or a part of the dark matter. Then an axion-like particle may provide the simple explanation for the observed line [4] or it may be the decay of a right neutrino [5]. Besides, an atomic dark matter can naturally explain the X-ray line also, without giving up on the WIMP paradigm, by identifying the low 3.55 keV energy scale with the hyperfine splitting of dark hydrogen [6]:

$$\Delta E = (8/3) (\alpha')^4 (m_e^2 m_p^2)/m_h^3 = 3.55 \text{ keV} \qquad (1)$$

Of course, it is a desired possibility. Here $m_h = m_e + m_p$ and $m_h, m_e, m_p$ are masses of dark hydrogen, dark electron and dark proton. Authors [6] suggest that the mass of dark hydrogen must be $m_h$~ 350-1300 GeV, the fine structure constant $\alpha'$ ~ 0.1 - 0.6, $m_p/m_e = 10^2 - 10^4$ to satisfy the eq. (1). Energy of annihilation of a dark electron and a dark positron is complicated to predict but it may be near 300 TeV in the case of two photons annihilation. For dark positronium (Ps) (Ps is an unstable lepton atom and it is a pure Coulomb system) our energetic estimates are the next: $L_\alpha^{Ps}$ ~3 GeV and $H_\alpha^{Ps}$ ~0.55 GeV. For the hyperfine ground state transition of dark Ps ($1^3S_1 \rightarrow 1^1S_0$) energy is E ~ 500 keV. We attempt to explain the excess in gamma diapason (1-3) GeV from the Galactic Center region by $L_\alpha$ line of a dark Ps. We are also predicting detection of $L_\alpha$ (E~6 GeV) and $H_\alpha$ (E~1.1 GeV) lines of dark hydrogen as well as the annihilation line of a dark positronium. Additional researches are probably necessary to detect these lines because of our estimates are very rough and speculative. An extremely interest presents the line of the hyperfine structure of dark Ps (~500 keV) and its annihilation line (~300 TeV). Lines of a dark Ps as well as a dark H may be a litmus paper for researches of atomic DM. Recently the nonabelian dark matter models for 3.5 keV X-rays were published in the article [7]. It was the first review on this theme.

**A laboratory positronium**

Positronium is some enigma in astrophysics - any lines of this simplest atomic system (electron + positron) are not observed till now. At first some worlds are necessary to say of "usual positronium" and it spectrum. A $L_\alpha$ line of Ps (2431Å) is impossible to observe in the direction of Galactic Center because of the strong extinction takes place here [8] although a positron excess was also observed in this direction [7]. An obvious signature of positrons is the 511 keV annihilation lines which were observed in at least two celestial objects: solar flares and the Galactic Center [9]. Another signature of positrons, which should allow improved resolution, is the recombination spectrum produced when a positron and an electron combine to form a positronium. The spectrum of positronium is analogous to the hydrogen recombination spectrum and frequencies can be calculated using the Rydberg formula. These frequencies are roughly a factor of two lower than the hydrogen recombination lines [9]. The average lifetime of Ps from singlet (para) and triplet (ortho) states is:

$$\tau_{para} \cong 1.25 \times 10^{-10} n^3 \text{ sec} \qquad \tau_{ortho} \cong 1.33 \times 10^{-7} n^3 \text{ sec} \qquad (2)$$

where: n is the main quantum number. The energies $E_n$ of the Ps levels are given by this formula (in eV): $\qquad E_n = -6.8/n^2 \qquad (3)$

The probabilities A of radiation transitions in Ps are two times smaller than the probabilities of transitions between similar states of hydrogen $A_{Ps} = (1/2) A_H$. The wavelengths of the emitted photons in Ps are longer than the analogous wavelengths of hydrogen $\lambda_{Ps} = 2 \lambda_H$. Theoretical calculations and experiments [10] imply that 90% of positrons annihilate after forming positronium. There are two competing processes which lead to the formation of Ps: (1) radiation recombination of free electrons and positrons: $e^+ + e^- = Ps$ and (2) charge exchange of positrons with neutral H: $e^+ + H = p + Ps$. If Ps is produced in an excited state then radiation decay through transitions to the ground state will be. Finally, Ps annihilates via 2 photon (511 keV) emission from the singlet S-state or via 3 photon (continuum 0 to 511 keV) emission from the triplet S-state. Probabilities of annihilation are [9, 11]:

$$A_{2\gamma} = (80/n^3) \times 10^8 \, sec^{-1} \qquad A_{3\gamma} = (0.075/n^3) \times 10^8 \, sec^{-1} \qquad (4)$$

Observations of ($e^+ e^-$) annihilation features have stimulated interest in the question about the possibility of the observations of the recombination lines of Ps. At temperatures less than $10^6$ K the rate of radiation capture in the process $e^+ + e^- \to$ $\to Ps + \gamma$ is higher than the rate of direct annihilation $e^+ + e^- \to 2\gamma$ [12]. Therefore, production of a lepton atom (Ps) must accompany every process of annihilation at "low temperatures". After entering in plasma an energetic positron will collide with electrons, protons, and ions. Because the probability of annihilation of a rapid positron is much less than the probability of energy loss, most of the initial positrons thermalizes and recombination occurs for thermal energies. The process of thermalization is strongly dependent on the physical state of the medium. In [8] we already analyzed the possibility of observation of the $L_\alpha$ – line of Ps ($\lambda \sim 2431$Å). As seen from results of that paper in a plasma at small T (T<10 eV) recombination takes place, in general, to the 2P levels (see Fig. in our articles [8, 13-14]). The production of the $L_\alpha$-line has a probability of 15% with respect to that of the annihilation line (in number of photons), i.e. $\beta = W_{2P \to 1S}/W_{2\gamma} \sim 0.15$ or $L_{2P \to 1S}/L_{2\gamma} \sim 1.5 \times 10^{-6}$. An only part of previous researches of Ps may be used to a dark Ps. In [13] we have shown that there is a power-law suppression of positronium formation in excited states relative to the ground state. The positronium $H_\alpha$ ($\lambda \sim 1.3$ μ) line, whose suppression factor is not so large ($\sim 1/n^3$), is the most suitable line for detection toward the Galactic Center for usual Ps. Knowing the flux in the annihilation line (511 keV), $\sim 10^{-3}$ photons $cm^{-2} sec^{-1}$, and suppression factor, we can obtain the expected flux in $H_\alpha$ line at Earth that a definite interest may present for us also:

$$F \sim 10^{-3}(1/27) \sim 3 \times 10^{-5} \, photons \, cm^{-2} sec^{-1} \approx 5 \times 10^{-17} erg \, cm^{-2} \, sec^{-1}$$

The spectral flux density in this line will be: $F_\nu \sim 5 \times 10^{-29}$ erg $cm^{-2} sec^{-1} Hz^{-1}$. The luminosity in any positronium line can be estimated by the formula:

$$L_{Ps} = h\nu \, \dot{n} \, (1/n^{\alpha+1}) \, \Phi \qquad (5)$$

where: $\dot{n}$ is the rate of e$^+$ formation by the positron's source (sec$^{-1}$), α=2 for low-lying levels, α=3 for n≫1, and Φ is the fraction of formed positronium atoms per positron. Thus, the Ps characteristic lines are the first lines of the Lyman L$_α$ (2431Å) n=2→n=1, Balmer H$_α$ (1.3 μ) n=3→n=2, Paschen P$_α$ (3.75 μ) n=4 →n=3, Brackett Br$_α$ (8.1 μ) n=5→n=4, Pfund Pf$_α$ (14.9 μ) n=6→n=5, Humphreys (24.7 μ) n=7→n=6 series, etc. Besides, in [14] we researched possibility of detecting the hyperfine ground state transition of positronum (λ~0.147 cm) under various astrophysical conditions. The maser effect may be also realizable. The expected fluxes F$_ν$ from this line in various regimes were estimated. They may be up to $10^{-1}$ mJy in the optimal case. This spin flip line was proposed in [15] for an interstellar radio communication since in mm diapason the galactic noise has a minimal value.

For estimates of dark Ps lines we proposed that the Coulomb interaction in the dark sector of the Universe is the same as the Coulomb interaction in the visible sector. An atomic dark matter will be a new intrigue ansatz of our study of space. The energies of strongest dark Ps lines are L$_α$ ~ 3 GeV, H$_α$ ~ 0.6 GeV; the hyperfine ground state line ($1^3S_1$→$1^1S_0$) has the energy near 500 keV; annihilation energy of a dark Ps may be near 300 TeV but our estimates are very rough.

## The Central Galactic Gamma Excess

Early studies have identified a spatially extended excess of ~1-3 GeV gamma rays from the region surrounding the Galactic Center, consistent with the emission expected from annihilating dark matter [16]. In this article authors confirm also that the angular distribution of the excess is approximately spherically symmetric and centered on the dynamical center of the Milky Way (within ~0.05$^0$ of Sgr A$^*$), showing no sign of elongation along the Galactic plane. The signal is observed to extend to at least ≃10$^0$ from the Galactic Center, disfavoring the possibility that this emission originates from millisecond pulsars. Several groups analyzing data from the Fermi Gamma-Ray Space Telescope have reported the detection of a gamma-ray signal from the inner few degrees around the Galactic Center (corresponding to a region several hundred parsecs in radius), with a spectrum and angular distribution compatible with that anticipated from annihilating dark matter particles [17-18]. While the spectrum and morphology of the Galactic Center and Inner Galaxy signals have been shown to be compatible with that predicted from the annihilations of an approximately 30-40 GeV WIMP annihilating to quarks (or a 7-10 GeV WIMP annihilating significantly to tau leptons)[19-20] other explanation may also be. Here we analyze the case when this excess may be produced by the $L_\alpha$ line of a dark positronium in a medium with the Compton scattering.

**Estimation of mass and dark Ps concentration in the Center of our Galaxy**.

The coefficient of radiation j and the observable flux density F on earth are defined by the next formulae:

$$j = (n\varepsilon A)/4\pi \qquad \text{erg/cm}^3\text{sec steradian} \qquad (6)$$

$$F = j\times\pi D\theta^2 = (n\varepsilon A/4) D\theta^2 \qquad \text{erg/cm}^2\text{sec} \qquad (7)$$

in which: n-concentration of Ps; $\varepsilon$-energy of $L_\alpha$ quant; $A\sim 10^8 \text{sec}^{-1}$ – probability of $L_\alpha$ radiation; D-diameter of radiation region; F-observable flux density in 3 GeV line; $\theta$-angle radius of the radiation region. If $\theta\sim 5^0$ (0.09 radian) then concentration of dark positronium is:

$$n = (4F/\varepsilon A D\theta^2) \approx 0.3\times 10^{-31} \qquad \text{cm}^{-3} \qquad (8)$$

and we can estimate the full mass of dark positronium in the Galactic Center. Since mass of dark Ps is near 300 TeV ($5\times 10^{-19}$ g) then according to (8) the density will be equal $5\times 10^{-19} \times 3\times 10^{-32} = 1.5\times 10^{-50}$ g/cm$^3$. In the region of size near 1 kpc the full mass of dark Ps will be near $1.5\times 10^{15}$ g. These values are very small, of course. The possibility to estimate the luminosity in an annihilation line of dark Ps (300 TeV) is appeared. It is L= $\varepsilon AnV\sim 5\times 10^{12} L_\odot$. It is an incredible value on our standard. A "dark flux" in this annihilation line on the Earth may reach the value F~$6\times 10^{-3}$ ph/cm$^2$sec.

**Evolution of $L_\alpha$ line of dark Ps for Compton Scattering**

We propose that a many times repeated the Compton effect at the line $L_\alpha$ of dark Ps (~3 GeV) in an usual plasma leads to redistribution photons on energy and this effect may form the observable extended gamma excess. As it was already shown in papers [21-22] a spectral line of the hard X-ray radiation is shifted down on energy and any line must be broaden. Of course, estimates may be done for $L_\alpha$ line of dark Ps in the center of our Galaxy. If $\varepsilon$ is the energy of radiated gamma quant then $\varepsilon_s$ is the energy of a scattered gamma quant. Since $\varepsilon \gg mc^2$ then electrons of usual plasma can be considered as resting ones and $\varepsilon_s$ will be [23]:

$$\varepsilon_s = \varepsilon/1 + [\varepsilon(1-\cos\theta)/mc^2] \leq \varepsilon \qquad (9)$$

In practical work for ε≫mc² a scattering takes place in the direction of a photon impulse in a narrow cone of scattering angles $\Delta\theta \approx \frac{mc^2}{\varepsilon} \ll 1$, therefore we have:

$$\varepsilon - (1/2)\, mc^2 \approx \varepsilon/1 + (\varepsilon/2mc^2)\,(mc^2/2)^2 \leq \varepsilon_s \leq \varepsilon \tag{10}$$

For estimates it necessary to put that for scattering an energy quant decreases on the value:
$$\Delta\varepsilon = \varepsilon - \varepsilon_s \approx (1/2)\, mc^2 \tag{11}$$

The cross-section of the scattering in our approximation is:

$$\sigma = \sigma_T \frac{3}{4x}\left(\ln x + \frac{1}{2}\right) = \sigma_T \frac{3mc^2}{8\varepsilon}\left(\ln\frac{2\varepsilon}{mc^2} + \frac{1}{2}\right) \tag{12}$$

here: $x = \frac{2\varepsilon}{mc^2}$ and $\sigma_T = (8/3)\pi\, r_e^2$ – cross-section of the Thomson scattering. A frequency of collisions of a photon with electrons is equal: ν=cσN where N is a concentration of electrons and then the reduction of the gamma quant energy is:

$$\frac{d\varepsilon}{dt} = -c\sigma N \Delta\varepsilon \approx -c\sigma N \frac{1}{2} mc^2 = -c\sigma_T N \frac{3}{16}(mc^2)^2 \frac{1}{\varepsilon}\left(\ln\frac{2\varepsilon}{mc^2} + \frac{1}{2}\right). \tag{13}$$

If l=ct is a size of the region in which escaping gamma radiation forms and τ = σ_T Nl is an optical depth on the Thomson scattering then the full reduction of gamma quant energy can reach the value:

$$\Delta\varepsilon \approx \tau \frac{3}{16}(mc^2)^2 \frac{1}{\varepsilon}\left(\ln\frac{2\varepsilon}{mc^2} + \frac{1}{2}\right). \tag{14}$$

A number of collisions during which a photon changes essentially its impulse direction is equal $\frac{\theta}{\Delta\theta} \approx \frac{1}{\Delta\theta} \approx \frac{\varepsilon}{mc^2} \gg 1$. Taking into account (11) a possible change of the quant energy which escapes from the scattering region can estimate:

$$\Delta\varepsilon = \frac{\theta}{\Delta\theta}\cdot\frac{1}{2}mc^2 \approx \frac{1}{2}\cdot\varepsilon \tag{15}$$

Substituting this expression in (14) we have:

$$\tau = \sigma_T\, Nl \approx \left(\frac{\varepsilon}{mc^2}\right)^2 \tag{16}$$

In our case ε=6000 mc$^2$ and then τ ~4x10$^7$. Obviously, mechanism of comptonization, suggested by us here does not work owing to the huge opt depth on the Thomson scattering. But this situation may be saved by anisotropy. If gamma radiation is centered in a narrow cone of pitch angles (Ψ) along the source radius, then the requests to the optical depth on the Thomson scattering are become by more soft in Ψ$^2$ time. Putting Ψ≤ (mc$^2$/ε) can to get τ≈1. Therefore, the observable width of this 3 GeV gamma line may be explained by processes of comptonization in usual plasma but in a very narrow solid angle. Other variant is more exotic. Let Ps dark atoms move at a speed near speed of light (v ~ c) then Δε ~ε (v/c) ~ε and we can explain the observable width of our line (~3GeV). These thermal speeds of Ps correspond to temperature of Ps dark atoms near 10$^{13}$K.

**Conclusion**

At first note that H$_α$ (λ~1.3 μ) line of usual Ps and its spin-flip line (λ~0.147 cm - orto-para transition) may be observed from our Galactic center with F$_v$ ~ 5x10$^{-29}$ erg cm$^{-2}$sec$^{-1}$ Hz$^{-1}$ (5μJy) and F$_v$ ~ 10$^{-27}$ erg cm$^{-2}$sec$^{-1}$ Hz$^{-1}$ (100 μJy) accordingly. Besides, the interpretation of 3.55 keV line in the frame of SM exits as yet [24]. It may be the L$_α$ line of hydrogen-like (π$^-$ $^3$H) mesoatoms. Note, that mesoatoms may be realized in magnetic columns of relativistic stars (binary pulsars) as we discussed this in [25].

The possibility of existence of atomic dark matter was also discussed by authors [26-27]. Authors [26] proposed that if dark matter is dominantly comprised of atomic bound states then formation of hydrogen-like dark atoms might take place in the early Universe. They proposed a dark sector charged under a hidden U (1) gauge symmetry and two species of fermions, a 'dark proton' and a 'dark electron', and that the dark matter abundance comes from a matter–anti-matter asymmetry. The existence of dark atoms implies that dark matter is coupled to dark radiation until the Universe cools beyond the binding energy of dark hydrogen. This has potentially interesting implications for structure formation because interactions in the dark sector may decouple much later than in a conventional CDM WIMP model. A dark recombination is other interesting moment discussed in [26]. Authors [27] note, that the DM problem points to the existence of new stable particles rather than modified gravity, and explore the hypothesis that the similar observed present-day densities for visible matter and DM suggest a common explanation for both. The visible and dark matter asymmetries may have arisen simultaneously through a first-order phase transition in the early Universe. The dark asymmetry can then be equal and opposite the usual visible matter asymmetry, leading to a Universe that is symmetric with respect to a generalized baryon number. If the dark matter is atomic as well as asymmetric then various cosmological and astrophysical constraints are derived very interesting [27]. Authors [3] note, that dark atom formation can occur efficiently in dense regions, such as the centers of a galactic halos. The formation of dark atoms is accompanied by emission of a dark photon, which can subsequently decay into Standard Model particles. Atomic character of DM is evident in the case of asymmetric DM coupled to a massless dark photon. DM remains atomic even if the dark U (1) (E-M) force is mildly broken. The important moment is here. In the case of a massless dark photon the gauge invariance prescribes that DM must be a multi-component system.

Our proposal to consider 3 GeV line in the direction of Galactic Center as a $L_\alpha$ line of positronium is speculative since the X-ray line (3.55 keV) may not be the line of hyperfine structure of the ground state of dark hydrogen. But if it is then other lines of dark hydrogen and dark positronium may be probably observable. For dark hydrogen these lines are $L_\alpha$ and $H_\alpha$ (E~ 6 GeV and 1.1 GeV accordingly). For dark positronium these lines are the line of annihilation (~300 TeV) with huge luminosity $L\sim 5\times 10^{12} L_\odot$ and the line of hyperfine structure (orto-para transition ~500 keV). The production of dark Ps is a separate complicated question. Flows of dark $e^+$ and dark $e^-$ are necessary in our picture. During flying they must draw together (Coulomb interaction) to produce a dark positronium before annihilation. That collimation is possible in jets of massive black holes. Relic black holes from dark components may evaporate dark $e^+$ and dark $e^-$ and production of Ps may also take place probably. This moment requests the careful investigation. Note, that large opt depths on Thomson scattering are typical for dense accretion discs around massive black holes. Therefore, in jets appearance of gamma-lines of dark Ps is not excluded the width of which will be defined by the process of comptonization which was considered by us.

The SM of physics of elementary particles is not a complete description of nature. New physics is certainly required to explain DM. In this a main scientific challenge of the modern time exists to us. The new science has to appear "spectroscopy of DM".